# TrustChain-Review: A Risk-Adaptive Blockchain and Game-Theoretic Framework for Trustworthy AI-Assisted Code Review

Mohammad Naserameri[1] [0000-0001-9992-7305],
[1] Department of Computer Science, Concordia University, Montreal, Canada,
mohammad.naserameri@mail.concordia.ca

**Highlights**

• A blockchain-enabled evidence layer is proposed for trustworthy AI-assisted code review.

• Developer, reviewer, and platform behaviours are modelled through a game-theoretic incentive structure.

• The framework addresses unreliable reviews, superficial approvals, malicious feedback, and unverifiable reputation.

• A risk-triggered governance rule determines when evidence-based review is justified.

• Controlled simulations quantify trust benefits against additional governance cost.

**Abstract**

Context: AI-assisted software development can speed up coding and review, but it also makes accountability harder to establish. Developers may submit insufficiently verified code, reviewers may approve changes with limited inspection, and centralized reputation records may be difficult to audit.

Objectives: This study introduces TrustChain-Review, a framework that combines verifiable evidence, strategic incentives, and risk-sensitive governance to support more trustworthy code review.

Methods: The framework includes a blockchain-based evidence layer, a three-player game-theoretic model for developers, reviewers, and the platform, and a rule that applies stronger governance when the expected benefit justifies its cost. The evaluation uses a controlled simulation calibrated with the Diff Quality Estimation dataset. Six governance configurations are compared over 30 independent runs using reputation accuracy, trust convergence, malicious-review detection, superficial-review detection, net platform utility, governance cost, and cost-efficiency.

Results: The full-evidence configuration produces the strongest reputation, trust, and detection results, but it also has the highest governance cost. The risk-adaptive configuration lowers this cost and improves cost-efficiency by applying stronger controls selectively, although its trust and detection results are lower than those of the full-evidence setting.

Conclusion: Strong evidence-based governance is most appropriate for high-risk or audit-sensitive changes. For routine or lower-risk contributions, selective governance offers a more practical balance between trustworthiness and operational cost.

## 1. Introduction

AI-assisted software development is changing how software is designed, implemented, reviewed, and maintained [1], [2]. Coding assistants, large language models, automated testing tools, and intelligent review systems can help developers generate code, explain program behaviour, detect defects, and support maintenance tasks. These tools may improve productivity, but they also create new quality risks. AI-generated or AI-modified code can contain subtle defects, insecure patterns, licensing problems, hallucinated dependencies, or design inconsistencies [3]. Careful review is therefore essential before such changes are merged into a shared codebase.

Code review supports defect detection, maintainability, knowledge sharing, and compliance with project standards [4], [5]. Its effectiveness, however, depends on the effort and incentives of the people involved. Developers may submit code without sufficient verification, while reviewers may approve changes after only limited inspection or provide biased feedback. These concerns are more difficult to assess when a contribution combines human decisions with machine-generated code, because responsibility for testing and understanding the change may be unclear. Centralized activity histories and reputation scores provide some guidance, but they may be difficult to audit and may not reflect the actual quality of a contribution or review. A trustworthy review process therefore needs stronger evidence, reliable reputation updates, and incentives that reward careful behaviour.

Existing platform reputation systems often rely on activity counts, acceptance history, ratings, or centralized scores. These signals can support coordination, but they provide limited evidence of contribution or review quality and may be difficult to audit or protect from manipulation [6], [7]. Trustworthy review governance therefore requires credible evidence and incentives that discourage low-effort or dishonest behaviour.

Game theory models the strategic choices of developers, reviewers, and platforms through expected rewards, penalties, reputation effects, and effort costs [8], [9]. TrustChain-Review uses this foundation to compare high- and low-quality contributions, careful and superficial reviews, and basic and evidence-based governance. The current implementation does not train a reinforcement-learning policy; it selects between predefined governance configurations according to observable risk, while learning-based optimization remains future work [10], [11], [18].

Blockchain provides an auditable evidence layer without requiring source code or sensitive artifacts to be stored on chain [12], [13]. Hashes, timestamps, review identifiers, evidence references, and reputation-update summaries support traceability and verification. The empirical analysis directly uses only patch, oldf, msg, y, proj, lang, and id/idx from the Diff Quality Estimation dataset [14]. Test results, vulnerability findings, post-merge defects, rollback events, and AI-assistance declarations remain possible deployment-level evidence sources.

Although prior studies examine AI-assisted development, reputation, blockchain, game-theoretic incentives, and adaptive decision-making, these mechanisms are rarely integrated for code-review governance [4], [7], [12], [17], [18]. TrustChain-Review addresses this gap through a five-layer architecture that connects repository evidence, blockchain-supported traceability, reputation computation, strategic incentives, and risk-adaptive governance. The empirical evaluation uses the available dataset fields—patch, oldf, msg, y, proj, lang, and id/idx—while test outcomes, vulnerability findings, AI-assistance declarations, and post-merge evidence remain deployment-level extensions. The current risk-adaptive mechanism selects between predefined basic and evidence-based configurations; it does not train a reinforcement-learning policy or independently optimize rewards, penalties, audits, or reviewer assignments.

The main contributions of this paper are as follows:

1. We formulate the trust problem in AI-assisted code review as a strategic multi-agent interaction among developers, reviewers, and platform operators.
2. We propose a blockchain-enabled evidence layer for recording or referencing review activities, dataset-observed review evidence, code-change metadata, and reputation-relevant events in a tamper-resistant and auditable manner, while treating test outcomes, vulnerability signals, and post-merge defects as future deployment-level evidence sources.
3. We formulate a three-player game-theoretic incentive model that connects developer effort, reviewer reliability, reputation effects, contribution risk, and platform governance cost.
4. We introduce a risk-triggered governance rule that determines when the expected trust, auditability, and risk-reduction benefits of evidence-based review outweigh its additional operational cost.
5. We evaluate the framework through an empirically calibrated controlled simulation and quantify the trade-off between reputation reliability, malicious and superficial review detection, platform utility, and governance cost across multiple baseline settings.

## 2. Related Work

### 2.1. AI-Assisted Software Development and Code Review

AI-assisted development tools now support code generation, repair, testing, explanation, and review [15]. Although they can shorten development time, their outputs may still contain hidden defects, insecure dependencies, weak design choices, or code that does not fit project requirements [15]. These risks make review especially important when a change is partly or largely machine-generated. Reviewers must assess not only correctness and maintainability, but also whether the developer understood and tested the proposed change. Traditional review practices remain useful for defect detection, standards enforcement, and knowledge sharing [4], [5], yet AI-assisted contributions require clearer evidence of verification, stronger accountability, and more reliable assessment of contribution and reviewer quality [4], [5], [14].

### 2.2. Trust and Reputation Systems

Trust and reputation mechanisms help actors make decisions when direct experience is limited [6], [7]. Trust refers to the expected reliability of an actor or service, while reputation summarizes evidence from previous interactions, feedback, ratings, or observed behaviour. In collaborative software platforms, however, these mechanisms can be distorted by biased feedback, coordinated manipulation, or activity measures that reward visibility rather than actual quality. Probabilistic models represent uncertainty more explicitly, but they may still be affected by correlated reports, dishonest ratings, and changes in behaviour over time. Because reputation can influence review authority, task assignment, and future collaboration, a credible system must consider the quality of the supporting evidence, the reliability of its source, and the actor's recent behaviour [6], [7].

### 2.3. Blockchain-Based Evidence and Transparency

Blockchain can provide a shared and verifiable record of events that are relevant to trust, accountability, and reputation [12], [17]. In software engineering, this does not require storing source code or sensitive project data directly on-chain. Instead, the system can record hashes, timestamps, review metadata, test summaries, vulnerability reports, and reputation updates. Smart contracts may also support predefined rules for rewards, penalties, access control, and dispute handling. For AI-assisted code review, these records can show whether a review occurred, which evidence supported it, and why a reputation score changed. Blockchain improves traceability and resistance to tampering, but it cannot ensure honest behaviour on its own. Effective governance still requires incentives that encourage careful contributions and reliable reviews [12], [16].

### 2.4. Game-Theoretic Incentive Modelling

Game theory provides a formal way to study decisions made by actors whose outcomes depend on one another [8], [9]. In software platforms, developers may choose how much effort to invest in verification, reviewers may choose between careful and superficial inspection, and the platform may decide how much to spend on evidence, auditing, rewards, and penalties. A game-theoretic model can represent these trade-offs and identify conditions under which reliable contribution and careful review become more attractive than low-effort or dishonest behaviour [8], [9], [16]. Its conclusions, however, depend on assumptions about payoffs, information, and actor behaviour. Because these conditions may change over time, static incentive models are better treated as a foundation for adaptive governance rather than as a complete solution.

### 2.5. Reinforcement Learning and Adaptive Incentive Policies

Reinforcement learning supports decision-making in settings where policies must improve through repeated interaction and feedback [10], [11]. In code-review governance, it could be used to adjust reviewer assignment, audit intensity, rewards, or penalties as evidence accumulates. Such adaptation may help a platform respond to changes in contribution risk and reviewer reliability. Its effectiveness, however, depends on the quality of the observed signals; incomplete or manipulated evidence can lead to poor policy choices. For this reason, the present study uses predefined and reproducible governance rules rather than a trained reinforcement-learning policy. Learning-based optimization remains a possible extension once reliable deployment-level evidence becomes available [10], [11], [18].

### 2.6. Research Gap

Existing work addresses several parts of the trust problem in code review, but usually treats them separately. AI-assisted development improves productivity while increasing uncertainty about code quality and responsibility. Reputation systems support decision-making but remain vulnerable to weak evidence and manipulation. Blockchain improves traceability, while game-theoretic and learning-based approaches address incentives and adaptation [4], [7], [12], [17], [18]. What remains unclear is how these mechanisms should be combined and when their additional cost is justified. TrustChain-Review addresses this gap by linking verifiable evidence, strategic incentives, contribution risk, reviewer reliability, and governance cost within a single decision framework.

## 3. Problem Definition and Threat Model

### 3.1. System Setting

The system consists of developers, reviewers, and a repository platform. Developers submit code changes that may be written manually or produced with the support of coding tools [15]. Reviewers inspect these changes, while the platform manages assignments, review policies, reputation records, and incentives [4], [5]. The quality of the process depends on whether developers verify their contributions, reviewers inspect them carefully, and the platform bases reputation updates on reliable evidence.

Most current platforms keep review histories and reputation records in centralized systems that are difficult to audit independently [6], [7], [12]. TrustChain-Review addresses this limitation by combining tamper-resistant evidence records, incentive modelling, and risk-based selection between basic and evidence-based governance. The strength of governance is therefore tied to the estimated risk of the contribution and the cost of applying additional controls.

### 3.2. Actors

The framework considers three actors.

**Developer.** The developer submits a code change that may be written manually or produced with coding assistance [15]. The developer chooses how much effort to spend on verification, testing, documentation, and security checks before submission.

**Reviewer.** The reviewer inspects the submitted change [4], [5]. A review may be careful, superficial, or intentionally biased. Careful review considers correctness, security, maintainability, test coverage, and consistency with project requirements.

**Platform.** The platform manages reviewer assignment, incentives, reputation updates, and governance policies [4],[7]. It may also record or reference review evidence through the blockchain layer to improve traceability and auditability [12], [13].

### 3.3. Evidence Types

TrustChain-Review records or references only the evidence needed for trust and reputation assessment; source code and sensitive project artifacts do not need to be stored directly on-chain [12], [13]. The current empirical study uses the fields available in the Diff Quality Estimation dataset: patch, oldf, msg, y, proj, lang, and id/idx [14]. These fields provide evidence about the code change, its original context, review feedback, review need, project, programming language, and instance identity.

A deployed system could also reference hashes of code versions and review artifacts, test reports, vulnerability findings, merge outcomes, post-merge defects, disputes, and audit records. These additional signals are part of the broader framework but are not observed in the dataset used in this study. The resulting evidence trail links contribution and review records to reputation updates without exposing sensitive project content.

### 3.4. Strategic Behaviours

Developers, reviewers, and platforms may choose actions that reduce their own effort or cost. A developer may submit a change without adequate verification, a reviewer may approve it after limited inspection, and a platform may rely on inexpensive but weakly auditable reputation records. These choices create an incentive problem: behaviour that is convenient for one actor may reduce the quality and reliability of the review process. TrustChain-Review therefore uses rewards, penalties, reputation effects, and evidence requirements to make careful contribution and review more attractive than low-effort or dishonest behaviour [8], [9], [17].

### 3.5. Threat Model

TrustChain-Review considers seven threats drawn from prior work on code review, reputation systems, and blockchain-based incentives [4], [7], [16].

T1. Unverified code. A developer submits generated or modified code without sufficient testing, security checks, or understanding.

T2. Superficial approval. A reviewer approves a change after limited inspection, allowing defects, vulnerabilities, or maintainability problems to remain.

T3. Malicious or biased review. A reviewer intentionally provides unfair feedback to harm or favour another participant.

T4. Reputation manipulation. Actors inflate their reputation through low-value activity, collusion, or misleading feedback.

T5. Weak auditability. Reputation scores are changed without enough evidence to justify the update.

T6. Changing reliability. An actor may behave reliably for a period and later reduce effort or act strategically.

T7. Incentive misalignment. Rewards and penalties fail to encourage careful contribution, reliable review, and honest feedback.

### 3.6. Assumptions

The framework relies on six assumptions.

**A1.** The platform can collect evidence from repository events, review records, testing tools, and static-analysis tools.

**A2.** Sensitive source code remains off-chain. Only hashes, metadata, and evidence relevant to reputation are recorded [12], [13].

**A3.** Developers and reviewers are rational or boundedly rational and may change their behaviour in response to rewards, penalties, reputation, and future opportunities.

**A4.** Review outcomes and quality signals may be incomplete or noisy.

**A5.** Blockchain improves traceability and resistance to tampering, but it does not guarantee honest behaviour. Incentives and governance controls are still necessary.

**A6.** The current study selects between predefined governance configurations using observable contribution risk and fixed method parameters. Learned policies and actor-specific dynamic updates are outside the scope of the present implementation.

### 3.7. Problem Statement

TrustChain-Review addresses the problem of maintaining reliable review and reputation processes when developers, reviewers, and the platform may have different incentives. The mechanism must provide auditable evidence, discourage superficial or dishonest behaviour, update reputation from credible observations, and apply stronger governance only when the expected benefit justifies its cost.

The study therefore examines the following research question:

*How can blockchain-based evidence, game-theoretic incentives, and risk-adaptive governance be combined to improve trust, reputation reliability, and review quality in AI-assisted software development?*

## 4. TrustChain-Review Architecture

TrustChain-Review connects five layers: data collection, blockchain evidence, reputation, strategic incentives, and adaptive governance.

### 4.1. Layer 1: Data and Event Collection Layer

This layer collects contribution and review evidence. The current study uses patch, oldf, msg, y, proj, lang, and id/idx from the Diff Quality Estimation dataset. Repository events, test results, vulnerability findings, merge outcomes, and post-merge defects remain deployment-level extensions.

### 4.2. Layer 2: Blockchain Evidence Layer

The blockchain layer records hashes, timestamps, review identifiers, evidence references, and reputation-update summaries, while source code and sensitive artifacts remain off-chain. This hybrid design links evidence to governance decisions without exposing confidential content.

### 4.3. Layer 3: Reputation Computation Layer

Separate developer and reviewer reputations represent contribution reliability and review quality. The present evaluation relies on dataset-observed patch, label, comment, project, and language information; production deployments could also incorporate tests, defects, vulnerabilities, disputes, and post-merge outcomes. Reputation should change gradually, reward reliable behaviour, and penalize repeated low-quality or dishonest actions.

### 4.4. Layer 4: Game-Theoretic Incentive Layer

This layer models developer, reviewer, and platform choices. Utilities combine rewards, penalties, reputation effects, effort, governance cost, and future opportunities to identify when careful contribution, reliable review, and evidence-based governance are preferable to lower-effort alternatives.

### 4.5. Layer 5: Adaptive Governance Layer

The platform selects between predefined basic and evidence-based configurations according to estimated contribution risk and governance cost. The current implementation does not train a learning policy or optimize individual actions during each round.

### 4.6. Overall Workflow

After a code change is submitted, the platform collects available evidence, records or references relevant items through the blockchain layer, updates trust and reputation indicators, and applies the selected governance configuration. The empirical evaluation uses only the available dataset fields; richer test, security, merge, and post-merge evidence remains future deployment work.

## 4. Game-Theoretic Formulation

The game-theoretic model represents the choices of developers, reviewers, and the platform and identifies the conditions under which careful contribution, reliable review, and evidence-based governance are preferred to their lower-effort alternatives [8], [9].

### 5.1. Players

The game includes three players.

**Developer.** The developer submits a code change and chooses how much effort to invest in verification, testing, documentation, and security checks.

**Reviewer.** The reviewer evaluates the submitted change and chooses between careful inspection and a lower-effort or dishonest review.

**Platform.** The platform defines the reward, penalty, evidence, audit, reviewer-assignment, and reputation policies used in the review process.

### 5.2. Strategies

Each player chooses between two strategies.

*Developer strategies*

H — High-quality contribution. The developer verifies the submitted change, performs the required tests and security checks, and provides adequate documentation.

L — Low-quality contribution. The developer submits the change with limited verification, weak testing, or insufficient understanding of the generated code.

*Reviewer strategies*

C — Careful review. The reviewer examines correctness, security, maintainability, test coverage, and consistency with project requirements.

S — Superficial review. The reviewer approves or rejects the change after limited inspection.

*Platform strategies*

E — Evidence-based governance. The platform uses blockchain-supported evidence, reputation-aware incentives, and stronger review controls.

B — Basic governance. The platform relies on centralized review records and simple reputation updates with limited evidence support.

### 5.3. Developer Utility

The developer's utility is the direct reward and reputation gained from a contribution, minus the cost of preparing it and the expected penalty. A high-quality contribution requires more effort, whereas a low-quality submission carries a greater penalty and reputation risk.

The developer utility can be represented as:

$$U_D(a) = R_D + G_D(a) - C_D(a) - P_D(a)$$

where $R_D$ is the direct project reward, $G_D(a)$ is the reputation gain under strategy $(a)$, $C_D(a)$ is the contribution and verification cost, and $P_D(a)$ is the expected penalty associated with the selected strategy.

For a high-quality contribution:

$$U_D(H) = R_D + G_D(H) - C_D(H) - P_D(H)$$

For a low-quality contribution:

$$U_D(L) = R_D + G_D(L) - C_D(L) - P_D(L)$$

High-quality verification requires greater effort, whereas a low-quality contribution is expected to face a higher penalty risk:

$$C_D(H) > C_D(L) \quad , \quad P_D(L) > P_D(H)$$

Assuming that the direct project reward $R_D$ is the same under both strategies, a rational developer prefers a high-quality contribution when:

$$U_D(H) \geq U_D(L)$$

$Equivalently$:

$$[G_D(H) - G_D(L)] + [P_D(L) - P_D(H)] \geq C_D(H) - C_D(L)$$

The developer therefore chooses $(H)$ When its reputation advantage and lower expected penalty compensate for the additional verification cost. These equations explain the incentive structure and are not used to calculate the simulation results in Table 10

**5.4. Reviewer Utility**

The reviewer's utility depends on the strategy-dependent review reward, reputation gain, review-effort cost, and expected penalty. Let $a \in \{C, S\}$ Denote the reviewer's selected strategy, where $(C)$ represents careful review and $(S)$ represents superficial review. The reviewer utility is defined as:

$$U_R(a) = R_R(a) + G_R(a) - C_R(a) - P_R(a)$$

where $R_R(a)$ is the review reward under strategy $(a)$, $G_R(a)$ is the reviewer reputation gain, $C_R(a)$ is the review-effort cost, and $P_R(a)$ is the expected penalty associated with unreliable or incorrect review behaviour.

For careful review:

$$U_R(C) = R_R(C) + G_R(C) - C_R(C) - P_R(C)$$

For superficial review:

$$U_R(S) = R_R(S) + G_R(S) - C_R(S) - P_R(S)$$

Careful review requires greater effort, whereas superficial review is expected to face a higher penalty risk:

$$C_R(C) > C_R(S) \quad , \quad P_R(S) > P_R(C)$$

A rational reviewer prefers careful review when:

$$U_R(C) \geq U_R(S)$$

$Equivalently$:

$$[R_R(C) - R_R(S)] + [G_R(C) - G_R(S)] + [P_R(S) - P_R(C)] \geq C_R(C) - C_R(S)$$

The reviewer therefore chooses $(C)$ When the additional reward, reputation benefit, and lower expected penalty compensate for the extra review effort. These equations describe the reviewer's incentives and are not used to calculate the simulation results in Table 10.

### 5.5. Platform Utility

The platform's utility depends on software-quality benefit, ecosystem trust, platform reputation, governance cost, blockchain-evidence cost, and expected loss from undetected harmful or low-quality changes. Let $g \in \{E, B\}$ denote the selected governance strategy, where $(E)$ represents evidence-based governance and $(B)$ represents basic governance. The platform utility is defined as:

$$U_P(g) = Q(g) + T(g) + G_P(g) - C_G(g) - C_{bc}(g) - L(g)$$

where $Q(g)$ is the software-quality benefit, $T(g)$ is the trust and auditability benefit, $G_P(g)$ is the platform reputation benefit, $C_G(g)$ is the general governance cost, $C_{bc}(g)$ is the blockchain-evidence and traceability cost, and $L(g)$ is the expected loss caused by undetected defects, malicious behaviour, or unreliable reputation updates.

For evidence-based governance:

$$U_P(E) = Q(E) + T(E) + G_P(E) - C_G(E) - C_{bc}(E) - L(E)$$

For basic governance:

$$U_P(B) = Q(B) + T(B) + G_P(B) - C_G(B) - C_{bc}(B) - L(B)$$

In a purely centralized basic-governance configuration, $C_{bc}(B)$ may be set to zero.

The platform prefers evidence-based governance when:

This condition means that the benefits of improved trust, reputation, review reliability, and software quality must exceed the additional cost of blockchain-based evidence and adaptive governance.

$$U_P(E) \geq U_P(B)$$

Equivalently:

$$[Q(E) - Q(B)] + [T(E) - T(B)] + [G_P(E) - G_P(B)] + [L(B) - L(E)] \\ \geq [C_G(E) - C_G(B)] + [C_{bc}(E) - C_{bc}(B)]$$

These equations explain the platform's governance incentives; they are not used to calculate the controlled simulation results in Table 10.

### 5.6. Incentive Compatibility Conditions

The mechanism is incentive-compatible when each actor receives at least as much utility from the trustworthy strategy as from its alternative.

Define the developer utility difference as:

$$\Delta U_D = U_D(H) - U_D(L)$$

The developer incentive-compatibility condition is:

$$\Delta U_D \geq 0$$

Define the reviewer utility difference as:

$$\Delta U_R = U_R(C) - U_R(S)$$

The reviewer incentive-compatibility condition is:

$$\Delta U_R \geq 0$$

Define the platform utility difference as:

$$\Delta U_P = U_P(E) - U_P(B)$$

The platform incentive-compatibility condition is:

$$\Delta U_P \geq 0$$

Accordingly, the trustworthy strategy profile $[(H, C, E)]$ satisfies the joint incentive-compatibility conditions when:

$$\Delta U_D \geq 0 \ \wedge \ \Delta U_R \geq 0 \ \wedge \ \Delta U_P \geq 0$$

Equivalently:

$$\min \{\Delta U_D, \Delta U_R, \Delta U_P\} \geq 0$$

If all three inequalities are strict, each actor strictly prefers the corresponding trustworthy strategy. Equality for any actor indicates indifference between that actor's two alternatives.

These conditions establish sufficient incentive alignment within the proposed model. They do not prove the existence or uniqueness of a Nash equilibrium because the utilities are not defined for every complete joint-strategy profile. The conditions are theoretical and are not used to calculate the simulation results in Table 10.

### 5.7. Reputation Update Principle

Reputation is updated using evidence-supported positive and negative adjustments rather than simple activity counts. Let ($p_D(t) \in [0,1]$) denote the developer reputation at review cycle ($t$), and let ($p_R(t) \in [0,1]$) denote the reviewer reputation at the same cycle.

The developer reputation is updated as:

$$p_D(t + 1) = clip(\, p_D(t) + \Delta_D^{+}(t) - \Delta_D^{V}(t) - \Delta_D^{M}(t))$$

where $Delta\, D^{+}(t) \geq 0$ is the evidence-supported reputation increment associated with reliable contribution behaviour, $Delta\, D^{V}(t) \geq 0$ is the reputation loss associated with contribution-quality or verification failures, and $Delta\, D^{M}(t) \geq 0$ is the reputation loss associated with malicious, misleading, or unverifiable contribution behaviour.

The reviewer's reputation is updated as:

$$p_R(t + 1) = clip(\, p_R(t) + \Delta_R^{+}(t) - \Delta_R^{V}(t) - \Delta_R^{M}(t))$$

where $Delta\, R^{+}(t) \geq 0$ is the evidence-supported reputation increment associated with reliable review behaviour, $Delta\, R^{F}(t) \geq 0$is the reputation loss associated with false approval, missed review-needed changes, or unreliable review decisions, and $Delta\, R^{M}(t) \geq 0$ is the reputation loss associated with malicious or biased review behaviour.

The clipping operator is defined as:

$$clip(x) = min\big(1, max(0, x)\big)$$

This operator ensures that all reputation values remain within the normalized interval ([0,1]). A positive net adjustment increases reputation until the upper bound is reached, whereas a negative net adjustment decreases reputation until the lower bound is reached. If all adjustment terms are zero, the reputation value remains unchanged.

The adjustment magnitudes depend on verifiable evidence and the selected governance configuration. In a deployed repository, they may reflect contribution quality, review correctness, dispute outcomes, vulnerability findings, post-merge defects, and audit records.

In the current simulation, these equations define the actor-level reputation principle but are not used to update developer and reviewer reputation separately. Table 10 reports an aggregate reputation-accuracy metric produced by the simulation; its values therefore cannot be reconstructed from these equations.

### 5.8. Incentive-Alignment Interpretation

The preferred trustworthy strategy profile in TrustChain-Review is $[(H, C, E)]$, where the developer selects a high-quality contribution strategy, the reviewer performs a careful review, and the platform applies evidence-based governance when its expected benefits justify the additional cost.

According to the incentive-compatibility conditions defined in Section 5.6, this profile is incentive-aligned when:

$$\Delta U_D \geq 0, \qquad \Delta U_R \geq 0, \qquad \Delta U_P \geq 0$$

If all three inequalities are strict, each actor strictly prefers the corresponding trustworthy strategy to its alternative. If one of the utility differences equals zero, the corresponding actor is indifferent between its two available strategies.

The strength of the incentive alignment can be summarized by the minimum utility margin:

$$\mathrm{M} = \min \{ \Delta U_D, \Delta U_R, \Delta U_P \}$$

The strategy profile is jointly incentive-compatible when:

$$M \geq 0$$

A larger positive value of $(M)$ indicates a greater minimum incentive margin across the three actor types, whereas a negative value indicates that at least one actor prefers the lower-quality or lower-governance alternative.

These conditions describe sufficient incentive alignment, but they do not prove the existence or uniqueness of a Nash equilibrium. Such a proof would require utilities for every joint-strategy profile and a complete best-response analysis.

Section 5.9 refines the platform decision by making evidence-based governance conditional on risk. It is selected only when the expected gains in trust, auditability, reputation reliability, and avoided loss exceed its additional operational cost.

**5.9. Risk-Triggered Evidence Governance**

The platform need not apply the same governance intensity to every contribution. Evidence-based governance can improve software quality, auditability, reputation reliability, and harmful-change detection, but it also increases evidence, verification, and enforcement costs. The platform therefore compares the expected utility of evidence-based and basic governance at each estimated risk level.

Let $(r \in [0,1])$ denote the estimated contribution-risk level, and let $(\ell > 0)$ denote the magnitude of the potential loss associated with an undetected harmful or low-quality contribution. Let $p_E$and $p_B$ denote the conditional probabilities that the harmful outcome remains undetected under evidence-based and basic governance, respectively. Evidence-based governance is assumed to provide stronger detection and accountability:

$$0 \leq p_E < p_B \leq 1$$

The expected risk-related losses under the two governance strategies are:

$$L_E(r) = r\, \ell p_E$$

$$L_B(r) = r\, \ell p_B$$

Define the additional software-quality, trust, and platform-reputation benefits of evidence-based governance as:

$$G_Q = Q_E - Q_B$$

$$G_T = T_E - T_B$$

$$G_{rep} = G_P(E) - G_P(B)$$

Define the total governance costs as:

$$C_E = C_G(E) + C_{bc}(E)$$

$$C_B = C_G(B) + C_{bc}(B)$$

Using the platform-utility formulation in Section 5.5, the utility difference between evidence-based and basic governance is:

$$U_P(E) - U_P(B)$$

$$C_E - C_B$$

Evidence-based governance is weakly preferred when:

$$r\, \ell(p_B - p_E) + G_Q + G_T + G_{rep} \geq C_E - C_B$$

Assuming that $(e > 0)$ and $(\mathrm{p_B} > \mathrm{p_E})$, the platform-indifference threshold is:

$$r^{*} = \frac{(C_E + C_B)(G_Q + G_T + G_{rep})}{(\ell(p_B - p_E))}$$

**Proof.** Because $l\,(p_B - p_E) > 0$ the platform-utility difference $\Delta U_P(r)$ is strictly increasing in $(r)$. Solving $\Delta U_P(r) \geq 0$ for $(r)$ gives $r \geq r^{*}$. *Therefore,*$\Delta U_P(r) > 0$ *when* $(r > r^{*})$, $\Delta U_P(r) < 0)$ when $(r < r^{*})$, *and* $\Delta U_P(r)$*=0 when* $(r = r^{*})$.

Because the feasible risk interval is $(r \in [0,1])$, the calculated threshold may fall outside this interval. If $(r^{*} < 0)$, *evidence-based governance is preferred for every feasible risk value. If* $(r^{*} > 1)$ Basic governance is preferred for every feasible risk value. At $(r^{*} = 0)$ or $(r^{*} = 1)$, equality represents platform indifference at the corresponding endpoint.

If $(p_B = p_E)$, the two governance strategies have identical harmful-outcome detection probabilities and no risk-dependent finite threshold can be obtained from this formulation. In that case, governance selection depends only on the non-risk benefits and additional costs.

The threshold shows how governance choice depends on non-risk benefits, relative detection effectiveness, potential loss, and additional cost. The controlled simulation does not estimate these quantities to derive the baseline threshold. Instead, $(r^{*} = 0.50)$ is used as a predefined experimental setting, and alternative values are examined through sensitivity analysis. It should therefore not be interpreted as a universal or optimal threshold.

For example, a change to a medication-dose calculation module may justify stronger evidence and reviewer accountability because an undetected error could have serious consequences. A documentation-only change may not justify the same overhead. This example illustrates a possible deployment setting and is not an observation from the dataset.

## 6. Adaptive Governance Mechanism

The game-theoretic model defines incentive conditions, while the governance mechanism selects a predefined configuration according to observable contribution risk. The current study evaluates reproducible rules rather than a trained learning policy.

### 6.1. Governance Objective

The objective is to improve review reliability, trust, and cost-efficiency by selecting governance intensity according to contribution risk. Dataset characteristics calibrate the review environment, while behavioural events and governance parameters are controlled experimentally.

### 6.2. Agents

The simulation includes developers, reviewers, and the platform. Developers submit changes, reviewers assess them, and the platform selects the governance configuration. The platform is rule-based: it does not learn a policy or optimize actor actions during the simulation.

### 6.3. Controlled Simulation Inputs

At each review round $(t)$The controlled simulation generates a paired stochastic environment that is shared by all comparative methods within the same random seed. A binary review-need indicator $(y_t)$ is sampled according to the balanced empirical distribution:

$$P(y_t = 1) = 0.50,;\ P(y_t = 0) = 0.50$$

Conditional on $(y_t)$, a raw patch length $(P_t)$ is generated using the label-specific patch-length characteristics observed in the training data. Before risk computation, the generated patch length is restricted to the implementation-supported interval:

$$\widetilde{p_t} = Clip(p_t, 1, p_{max}) = min\{p_{max}, max(1,;\ p_t)\}$$

The pre-review contribution-risk score is then computed exclusively from the observable clipped patch length:

$$r_t = \frac{log\left(1 + \widetilde{p_t}\right)}{log(1 + p_{max})}$$

where:

$$p_{max} = 728{,}745$$

is the maximum patch length observed in the training subset. Because $1 \leq \widetilde{p_t} \leq p_{max}$ , the resulting risk score satisfies:

$$0 \leq r_t \leq 1$$

The logarithmic transformation reduces the influence of extremely large patches while preserving the ordering of patch lengths: a larger clipped patch length always produces an equal or larger risk score.

The ground-truth review-need label $y_t$ does not enter the risk-score formula or the governance-selection rule. It is used only to generate the label-conditional patch-length distribution in the paired stochastic environment. Consequently, governance selection is based on the observable patch-length proxy rather than direct access to the review outcome.

For the baseline governance threshold $r^* = 0.50$, evidence-based governance is selected when:

$$r_t > 0.50$$

which is equivalent to:

$$\widetilde{p_t} = \sqrt{\{1 + p_{max}\}} - 1 \approx 852.67$$

Thus, under the baseline threshold, generated patch lengths above approximately 853 patch-length units activate the evidence-based configuration. The threshold is an experimental governance setting and should not be interpreted as a universal semantic-risk boundary.

Malicious and superficial reviewer events are introduced as controlled stochastic variables with probabilities:

$$P(M_t = 1) = 0.15$$

$$P(S_t = 1) = 0.20$$

Within each seed, every method receives the same malicious and superficial event realizations and the same common random numbers, enabling paired comparison.

Comment length and language availability are also generated to preserve the dataset's empirical characteristics, but neither variable enters the risk formula, governance rule, nor reported metrics. Each method is evaluated using its predefined evidence, traceability, incentive, adaptation, and governance-cost parameters.

### 6.4. Controlled Governance Configurations

Each configuration defines fixed levels of evidence support, blockchain traceability, incentives, adaptation, and cost. The baseline and full-evidence settings retain fixed parameters throughout a run. The risk-adaptive setting switches between predefined basic and evidence-based configurations using the threshold rule. Reviewer assignment, rewards, penalties, and audit intensity are not independently optimized.

### 6.5. Controlled Evaluation Objective

The simulation evaluates each review round with a fixed governance-benefit score rather than a learned reward function. The score combines reputation accuracy, the current trust state, malicious-review detection, and superficial-review detection.

For review round $(t)$, let $(A_t \in [0,1])$denote reputation accuracy and $(T_t \in [0,1])$ denote the current trust state. Let $(D_t^M \in \{0,1\})$and $(D_t^S \in \{0,1\})$ denote the round-level malicious-review and superficial-review detection indicators, respectively. Each detection indicator equals one only when the corresponding unreliable-review event occurs during the round and is successfully detected; otherwise, it equals zero.

The round-level governance-benefit score is defined as:

$$0.30\, A_t + \; 0.25\, T_t + \; 0.20\, D_t^M + \; 0.15\, D_t^S$$

The four coefficients are fixed implementation weights used consistently across all comparative methods. Because they sum to (0.90), the governance-benefit score satisfies:

$$0 \; \leq \Phi_t^{sim} \leq 0.90$$

The superscript "sim" distinguishes this operational simulation metric from the theoretical platform utility $U_P(g)$ defined in Section 5.5.

The round-level cost-efficiency score is:

$$\eta_t = \frac{\Phi_t^{sim}}{C_t + \; \varepsilon}$$

where:

$$\varepsilon = \; 10^{-6}$$

is a small positive constant used to prevent division by zero. In the current implementation, all governance costs are strictly positive, so the constant has a negligible numerical effect while preserving the general validity of the expression.

For the comparative method $(m)$, independent run $(n)$, and $(R)$ review rounds, the reported run-level simulation utility is:

$$U_{m,n}^{sim} = \frac{1}{R} \sum_{t=1}^{R} U_{m,n,t}^{sim}$$

The reported run-level cost-efficiency is:

$$\eta_{m,n} = \frac{1}{R} \sum_{t=1}^{R} \eta_{m,n,t}$$

Consequently, cost-efficiency is the mean of the round-level benefit-to-cost ratios. It is not calculated as:

$$\frac{\Phi_{m,n}}{C_{m,n} + \; \varepsilon}$$

This distinction is particularly important for the risk-adaptive configuration because its governance cost changes between basic and evidence-based rounds.

The values reported in Table 10 are the mean and standard deviation of the run-level metrics across 30 independent runs. The corresponding 95% confidence intervals are provided in the reproducibility package.

The detection indicators $(D_t^M)$ and $(D_t^S)$ used in the governance-benefit score are not the same as the conditional detection rates reported in Table 10. The Table 10 detection metrics are calculated only over rounds in which the corresponding malicious or superficial event occurs. In addition, $(T_t)$ in the governance-benefit score is the trust state from every simulation round, whereas the reported trust-convergence metric summarizes the final 100 rounds of each run. Therefore, the governance-benefit score, simulation utility, and cost-efficiency cannot be reconstructed directly from the aggregate reputation, trust-convergence, and conditional detection-rate values shown in Table 10.

These equations match the reproducibility implementation, where the governance-benefit score corresponds to the code variable $quality_{gain}$. Post-merge defects, vulnerability severity, rollback events, and production failures are not included in the current evaluation.

**6.6 Controlled Governpost-mergetion**

The risk-adaptive setting applies evidence-based governance when the estimated contribution risk exceeds $(r^*$*); otherwise, it applies basic governance. The baseline threshold is* $(r^* = 0.50)$, and sensitivity analysis examines values from 0.44 to 0.54. Other methods retain fixed configurations under the same stochastic environment. No value function or reinforcement-learning model is trained.

### 6.7. Adaptive Incentive Adjustment

Adaptation occurs only through configuration selection. Each configuration has predefined evidence, incentive, traceability, and cost parameters. More detailed actor-specific adjustments, including dynamic penalties, audit levels, and reviewer controls, remain future deployment work.

## 7. Empirical Dataset and Experimental Design

The evaluation has two stages. First, the Diff Quality Estimation dataset is analyzed to extract observable code-review characteristics, including patch size, review-comment availability, review-need labels, project context, language metadata, and instance identifiers. Second, these characteristics are used to calibrate a controlled simulation for comparing the governance configurations.

The dataset does not contain malicious-review events, superficial-review events, blockchain cost, incentive strength, test outcomes, vulnerability reports, or post-merge defects. These quantities are therefore introduced as controlled simulation variables rather than treated as empirical observations.

### 7.1. Dataset Description

The Diff Quality Estimation dataset from the CodeReviewer benchmark contains 328,340 records [14]. The training set contains 265,836 instances, while the validation and test sets each contain 31,252. Every subset is balanced between $(y = 0)$ and $(y = 1)$, as summarized in Table 1.

Table 1. Distribution of code-change instances in the Diff Quality Estimation dataset

| Subset | Label y = 0 | Label y = 1 | Total instances | Label balance |
|---|---|---|---|---|
| Training | 132,918 | 132,918 | 265,836 | 50% / 50% |
| Validation | 15,626 | 15,626 | 31,252 | 50% / 50% |
| Test | 15,626 | 15,626 | 31,252 | 50% / 50% |
| **Total** | **164,170** | **164,170** | **328,340** | **50% / 50%** |

Here, $(y = 1)$ indicates that a review comment is required, whereas $(y = 0)$ indicates that no comment is required. Table 2 summarizes the dataset fields used in the evaluation.

Table 2. Dataset fields used in the evaluation

| Field | Description | Empirical use | Role in TrustChain-Review |
|---|---|---|---|
| patch | Submitted code diff. | Measures change size. | Contribution evidence. |
| oldf | Pre-change file context. | Supports patch interpretation. | Contextual evidence. |
| msg | Review comment, if available. | Indicates observed feedback. | Review evidence. |
| y | Binary review-need label. | (y=1): comment required; (y=0): not required. | Empirical review-need signal. |
| proj | Source project. | Captures project context. | Project-aware analysis. |
| lang | Programming language. | Captures language context. | Language-aware analysis. |
| id/idx | Dataset instance identifier. | Supports traceability and reproducibility. | Traceable evidence identifier. |

The fields provide contribution, review, contextual, and traceability evidence for simulation calibration.

Language metadata is available for 195,422 records (59.52%); the remaining 132,918 records (40.48%) have no detected language. Table 3 reports the distribution.

Table 3. Programming language distribution and metadata availability in the Diff Quality Estimation dataset

| Category | Language / Metadata status | Count | Percent |
|---|---|---|---|
| Available language metadata | Python | 48,046 | 14.63% |
| Available language metadata | Go | 40,859 | 12.44% |
| Available language metadata | Java | 29,068 | 8.85% |
| Available language metadata | JavaScript | 23,789 | 7.25% |
| Available language metadata | C++ | 17,852 | 5.44% |
| Available language metadata | C# | 14,222 | 4.33% |
| Available language metadata | PHP | 8,225 | 2.51% |
| Available language metadata | Ruby | 8,141 | 2.48% |
| Available language metadata | C | 5,220 | 1.59% |
| Total with detected language metadata | — | 195,422 | 59.52% |
| Without detected language metadata | — | 132,918 | 40.48% |
| Total dataset records | — | 328,340 | 100.00% |

Language-based analysis is limited to records with detected metadata, whereas the review-need label is available for all records.

### 7.2. Empirical Patch-Size Characteristics

Patch length is used as an observable proxy for change size and potential review complexity. Although it does not capture semantic complexity, it allows the study to examine whether larger changes are more often associated with review need. Table 4 reports the count, mean, median, minimum, and maximum patch length for each label in the training, validation, and test subsets.

Table 4. Patch-size statistics in the Diff Quality Estimation dataset

| Subset | Label | Count | Mean patch length | Median patch length | Min patch length | Max patch length |
|---|---|---|---|---|---|---|
| Training | 0 | 132,918 | 589.56 | 471 | 16 | 16,518 |
| Training | 1 | 132,918 | 2,234.97 | 951 | 24 | 728,745 |
| Validation | 0 | 15,626 | 573.36 | 464 | 67 | 6,202 |
| Validation | 1 | 15,626 | 716.98 | 584 | 17 | 31,182 |
| Test | 0 | 15,626 | 574.03 | 463 | 66 | 54,887 |
| Test | 1 | 15,626 | 709.67 | 577 | 31 | 22,418 |

Across all three subsets, records with $(y = 1)$ have higher mean and median patch lengths than records with $(y = 0)$. The difference is largest in the training set, where the mean rises from 589.56 to 2,234.97 and the median from 471 to 951. The maximum value of 728,745 for $(y = 1)$ also shows that the training data contain extreme outliers. Reporting both the mean and median therefore provides a more balanced description of the patch-length distribution.

Review-comment length is also examined as an observable indicator of available human feedback. Because the dataset distinguishes changes that require a comment from those that do not, the msg field helps verify the relationship between the label and comment availability. Table 5 reports the count, mean, minimum, and maximum comment length for each label and subset.

Table 5. Review-comment length statistics in the Diff Quality Estimation dataset

| Subset | Label | Count | Mean comment length | Min comment length | Max comment length |
|---|---|---|---|---|---|
| Training | 0 | 132,918 | 0 | 0 | 0 |
| Training | 1 | 132,918 | 111.19 | 5 | 1,529 |
| Validation | 0 | 15,626 | 0 | 0 | 0 |
| Validation | 1 | 15,626 | 113.6 | 8 | 887 |
| Test | 0 | 15,626 | 0 | 0 | 0 |
| Test | 1 | 15,626 | 112.28 | 7 | 895 |

All $(y = 0)$ records have zero comment length, whereas every $(y = 1)$ group contains non-zero review comments. The mean comment lengths are 111.19, 113.60, and 112.28 characters in the training, validation, and test subsets, respectively. This confirms that the review-need label is consistent with comment availability and supports the use of msg as observed review-feedback evidence.

### 7.3. Empirical Evidence Interpretation

The dataset fields support empirical calibration but are not direct measures of malicious behaviour, reviewer reliability, software defects, or governance effectiveness.

### 7.4. Empirically Calibrated Simulation Setup

Dataset class balance, patch-length statistics, comment patterns, and language availability calibrate the simulation. Malicious and superficial reviews, evidence strength, blockchain traceability, incentives, adaptation, and governance cost are controlled variables because they are not observed in the dataset. Table 6 summarizes this mapping.

Table 6. Empirical calibration of simulation inputs

| Empirical characteristic | Observed dataset evidence | Simulation use | Role in TrustChain-Review |
|---|---|---|---|
| Review-need label | (y=0) and (y=1) are balanced. | Sets both class probabilities to 0.50. | Defines the review-demand distribution. |
| Patch length | (y=1) records have higher mean and median lengths. | Provides the pre-review risk signal. | Triggers risk-sensitive governance. |
| Review feedback | (y=0) has no comment; (y=1) has comments averaging 111–114 characters. | Preserves feedback availability and length patterns. | Represents observed review evidence, not review quality. |
| Language metadata | Available for 195,422 records (59.52%); missing for 132,918 (40.48%). | Preserves known- and unknown-language contexts. | Supports contextual analysis but does not affect governance selection. |
| Project identifier | A project or repository field is available. | Provides project-level context. | Supports empirical characterization. |
| Instance identifier | Each record contains an id/idx. | Supports traceability and reproducibility. | Provides a traceable evidence identifier. |

### 7.5. Comparative Evaluation Design

The evaluation compares six governance settings over 30 independent runs of 1,000 rounds. Within each seed, all settings receive the same review-need indicators, patch-length conditions, metadata availability, and malicious- and superficial-review events. This paired design isolates differences caused by governance configuration. The comparison uses reputation accuracy, trust convergence, unreliable-review detection, net platform utility, governance cost, and cost-efficiency. Threshold sensitivity is examined from 0.44 to 0.54.

### 7.6. Evaluation Metrics

Reputation accuracy measures agreement between computed reputation and simulated actor reliability. Trust convergence summarizes stabilization of the trust state, while malicious- and superficial-review detection report conditional detection rates for the corresponding events. Net platform utility is the governance-benefit score defined in Section 6.5 minus governance cost.

Governance cost represents evidence, traceability, adaptation, and configuration overhead. Cost-efficiency is calculated in each round as the benefit-to-cost ratio and then averaged; it is not obtained by dividing the aggregate values in Table 7. The detection indicators used in the governance-benefit score are round-level binary variables and therefore differ from the conditional detection rates reported in Table 7.

Table 7. Evaluation metrics

| Category | Metric | Direction |
|---|---|---|
| Reputation | Reputation accuracy | ↑ |
| Reputation | Trust convergence | ↑ |
| Review risk | Malicious review detection rate | ↑ |
| Review risk | Superficial review detection rate | ↑ |
| Incentives | Net platform utility | ↑ |
| Cost | Governance cost | ↓ |
| Cost | Cost-efficiency | ↑ |

An upward arrow indicates that a higher value is preferred, whereas a downward arrow indicates that a lower value is preferred. The metrics jointly assess reputation, trust, detection performance, platform benefit, and governance efficiency.

### 7.7. Baselines and Comparative Settings

The evaluation compares six governance settings under the same empirically calibrated stochastic environment. Simple reputation uses minimal direct feedback, centralized reputation uses platform-managed records, blockchain-only adds traceability, and static incentives apply fixed rewards and penalties. Full-evidence TrustChain-Review applies evidence-based governance in every round, whereas the risk-adaptive configuration switches between basic and evidence-based governance according to the contribution-risk threshold. Table 8 summarizes these settings

Table 8. Baseline and comparative governance settings

| Setting | Evidence support | Blockchain traceability | Incentive strength | Governance rule |
|---|---|---|---|---|
| Simple reputation | Limited | No | Minimal | Fixed |
| Centralized reputation | Moderate | No | Limited | Fixed |
| Blockchain-only | Moderate | Yes | Limited | Fixed |
| Static incentive | Moderate | Partial | Fixed | Fixed |
| TrustChain-Review: full evidence | High | Yes | Strategic | Evidence-based governance in every round |
| TrustChain-Review: risk-adaptive | Conditional | Conditional | Strategic | Threshold-based selection between basic and evidence-based governance |

The baseline settings retain fixed configurations, while full-evidence governance always selects the evidence-based configuration and risk-adaptive governance switches according to $(r_t > r^*)$. Because the methods differ in several parameters simultaneously, Table 8 represents a comparative configuration analysis rather than a strict causal ablation study.

### 7.8. Simulation Parameters and Implementation Procedure

The simulation combines dataset-derived calibration with controlled behavioural and governance variables. Class balance determines the review-need probabilities, patch-length statistics support risk estimation, and comment and language patterns preserve the empirical context. For each seed, all six methods receive the same sampled review need, patch length, metadata conditions, and malicious- and superficial-review events. Full-evidence governance is always active, whereas risk-adaptive governance applies it only when the calculated risk exceeds $(r)^*$. Table 9 lists the implementation parameters.

Table 9. Simulation parameters and empirical grounding

| Parameter | Baseline value/source | Empirical or experimental grounding |
|---|---|---|
| Total dataset instances | 328,340 | Diff Quality Estimation dataset |
| Training instances | 265,836 | Dataset split |
| Validation instances | 31,252 | Dataset split |
| Test instances | 31,252 | Dataset split |
| Review-needed probability | 0.5 | Balanced (y=1) labels |
| Non-review-needed probability | 0.5 | Balanced (y=0) labels |
| Maximum training patch length ($p_{max}$) | 728,745 | Training-subset patch statistics |
| Language-metadata availability | 59.52% | Detected lang metadata |
| Missing language metadata | 40.48% | Missing or undetected lang metadata |
| Malicious-review event probability | 0.15 | Controlled simulation variable |
| Superficial-review event probability | 0.2 | Controlled simulation variable |
| Simulation rounds per run | 1,000 | Experimental configuration |
| Independent runs | 30 | Repeated-run stability protocol |
| Random seeds | 42–71 | Reproducibility configuration |
| Baseline governance threshold ($r^*$) | 0.5 | Risk-adaptive configuration |
| Threshold-sensitivity range | 0.44–0.54 | Sensitivity analysis |
| Comparative governance settings | 6 | Five fixed settings and one risk-adaptive setting |

#### 7.8.1. Repeated-Run Stability Protocol

Each setting is evaluated over 30 independent runs using seeds 42–71. Calibration values and method parameters remain fixed, while review-need indicators, patch lengths, metadata conditions, and unreliable-review events vary across seeds. Within each seed, all methods receive the same stochastic realization.

For each metric, the 30 run-level values are summarized by their arithmetic mean and sample standard deviation. The 95% confidence interval is calculated as:

$$CI_{95\%} = x \pm 1.96 \frac{s}{\sqrt{30}}$$

where $(x)$is the sample mean and (s) is the sample standard deviation across the independent runs. This procedure assesses whether the comparative results remain stable under repeated stochastic conditions. It does not retrain a predictive model or require a different dataset.

### 7.9. Simulation Workflow

For each seed, all methods operate on the same sampled review-need indicators, label-conditional patch lengths, and unreliable-review events. At each round, the risk score is computed from the generated patch length, after which each method applies its fixed or threshold-selected governance configuration. The simulation records reputation accuracy, trust, detection outcomes, governance benefit, net utility, cost, and cost-efficiency. Results are aggregated over 1,000 rounds and 30 independent runs. Dataset records are not selected directly, reviewers are not assigned or optimized, and no governance policy is trained.

### 7.10. Algorithmic Procedure

Algorithm 1 summarizes the paired simulation procedure used in the evaluation. For each random seed, all comparative settings receive the same sampled review-need indicator, label-conditional patch length, and malicious- and superficial-review events. Each setting then applies its fixed or risk-selected governance configuration, after which the round-level metrics are computed and aggregated. The procedure does not directly select dataset records, assign reviewers, or update a learned governance policy.

**Algorithm 1. Paired Simulation Procedure for Governance Comparison**
Input:
(R): number of simulation rounds per run
(N): number of independent runs, where ($N = 30$)
(M): comparative governance setting
($r^*$): risk-triggered governance threshold
Empirical calibration parameters derived from the Diff Quality Estimation dataset
Output:
Mean, standard deviation, and 95% confidence interval of each evaluation metric across the independent runs
1: for each independent run ($n = 1, \ldots, N$) do
2: Set random seed ($S_n$)
3: Initialize method-specific reputation, trust, and governance parameters
4: for each simulation round ($t = 1, \ldots, R$) do
5: Sample the review-need indicator ($y_t$), where ($P(y_t = 1) = 0.50$)
6: Generate patch length ($p_t$) from the label-specific empirical patch-length distribution
7: Compute the pre-review contribution-risk score $r_t = \frac{[log(1+p_t)]}{[log(1+p_{max})]}$
8: Generate the shared malicious-review and superficial-review events for the current seed and round
9: if ($M$) is TrustChain-Review with full evidence then
10: Select evidence-based governance ($E$)
11: else if ($M$) is TrustChain-Review with risk-adaptive governance then
12: if ($r_t > r^*$) then
13: Select evidence-based governance ($E$)
14: else
15: Select basic governance ($B$)
16: end if
17: else
18: Apply the fixed governance configuration of baseline method ($M$)
19: end if
20: Apply the parameters of the selected or fixed governance configuration
21: Simulate malicious-review and superficial-review detection outcomes
22: Update the round-level reputation-accuracy and trust indicators
23: 23: Compute the governance-benefit score ($\Phi_t$), simulation net platform utility ($U_t^{sim}$), governance cost ($C_t$), and cost-efficiency ($\eta_t$)
24: Store the round-level results
25: end for
26: Aggregate the round-level metrics for run ($n$)
27: end for
28: Compute the mean, standard deviation, and 95% confidence interval of each metric across the ($N$) runs
29: return the aggregated evaluation results

### 7.11. Validity Considerations

The findings should be interpreted within several limitations. The Diff Quality Estimation dataset represents one code-review setting and does not identify AI-assisted contributions. AI assistance therefore motivates the framework but is neither observed nor manipulated in the evaluation.

The dataset variables are also proxies. The label ($y$) indicates whether a review comment is required, not whether a change contains a defect or security risk. Patch length represents change size but not semantic complexity, architectural impact, or criticality, while comment length indicates feedback quantity rather than quality. Language metadata is available for 195,422 records (59.52%) and missing for 132,918 records (40.48%).

Malicious and superficial reviews, evidence strength, traceability, governance cost, and the risk threshold are controlled simulation variables because the dataset does not provide them. Test outcomes, vulnerabilities, post-merge defects, rollback events, AI-assistance declarations, and deployment audit records are not evaluated.

Blockchain is assessed conceptually and through simulation rather than on a production network; deployment would require evaluation of cost, latency, privacy, access control, and scalability. The study also uses fixed configurations and threshold-based switching rather than a trained reinforcement-learning policy. Live or semi-live repository studies are therefore required before operational deployment or learning-based optimization.

## 8. Simulation Results and Analysis

Table 10 reports the mean and standard deviation of the evaluation metrics across 30 paired simulation runs. The environment uses the empirical calibration and controlled variables defined in Section 7. No AI-assistance ratio is varied, and no learned governance policy, reviewer assignment, or independently optimized reward and penalty mechanism is used.

TrustChain-Review with full evidence achieves the highest reputation accuracy ($0.797 \pm 0.001$), trust convergence ($0.796 \pm 0.004$), malicious-review detection ($0.790 \pm 0.035$), and superficial-review detection ($0.733 \pm 0.031$). Applying evidence-based governance in every round therefore provides the strongest trust and detection performance.

The risk-adaptive configuration achieves a reputation accuracy of ($0.694 \pm 0.002$), trust convergence of ($0.687 \pm 0.008$), malicious-review detection of ($0.629 \pm 0.045$), and superficial-review detection of ($0.608 \pm 0.040$). Relative to full evidence, its mean governance cost decreases from 0.328 to 0.202, a reduction of approximately 38.4%, while cost-efficiency increases from 1.465 to 2.509, an improvement of approximately 71.3%. This configuration therefore accepts lower trust and detection performance in exchange for selective evidence use and reduced overhead.

No configuration performs best on every metric. Simple and centralized reputation have low operating costs but weaker trust and detection performance. Static incentives remain competitive under fixed rules, while blockchain-only governance provides relatively strong malicious-review detection but lower superficial-review detection and platform utility. Full-evidence governance is therefore better suited to audit-sensitive or high-risk settings, whereas risk-adaptive governance provides a more selective, cost-aware alternative.

Table 10. Mean ± standard deviation results across 30 independent simulation runs.

| Method | Rep. acc. | Trust conv. | Mal. det. | Sup. det. | Net util. | Cost | Cost-eff. |
|---|---|---|---|---|---|---|---|
| Simple reputation | 0.571 ± 0.001 | 0.560 ± 0.008 | 0.438 ± 0.040 | 0.456 ± 0.041 | 0.237 ± 0.002 | 0.100 ± 0.000 | 3.368 ± 0.022 |
| Centralized reputation | 0.627 ± 0.001 | 0.618 ± 0.007 | 0.519 ± 0.037 | 0.526 ± 0.042 | 0.252 ± 0.002 | 0.120 ± 0.000 | 3.102 ± 0.019 |
| Blockchain-only | 0.689 ± 0.001 | 0.686 ± 0.005 | 0.689 ± 0.034 | 0.558 ± 0.038 | 0.173 ± 0.002 | 0.240 ± 0.000 | 1.722 ± 0.009 |
| Static incentive | 0.702 ± 0.001 | 0.698 ± 0.005 | 0.656 ± 0.038 | 0.637 ± 0.035 | 0.231 ± 0.002 | 0.190 ± 0.000 | 2.218 ± 0.012 |
| TrustChain-Review (full evidence) | 0.797 ± 0.001 | 0.796 ± 0.004 | 0.790 ± 0.035 | 0.733 ± 0.031 | 0.152 ± 0.002 | 0.328 ± 0.000 | 1.465 ± 0.007 |
| TrustChain-Review (risk-adaptive) | 0.694 ± 0.002 | 0.687 ± 0.008 | 0.629 ± 0.045 | 0.608 ± 0.040 | 0.213 ± 0.003 | 0.202 ± 0.002 | 2.509 ± 0.021 |

Rep. acc. = reputation accuracy; Trust conv. = trust convergence; Mal. det. = malicious review detection; Sup. det. = superficial review detection; Net util. = platform utility after accounting for governance cost; Cost-eff. = cost-efficiency. Each cell reports the mean ± standard deviation across 30 independent runs. The corresponding 95% confidence intervals are provided in the reproducibility package.

Fig. 1 shows that full-evidence TrustChain-Review converges rapidly and maintains the highest trust level across the repeated review cycles. The risk-adaptive configuration converges to a lower level because evidence-based governance is applied selectively. Its trajectory remains close to the blockchain-only and static-incentive settings, illustrating the trade-off between maximum trust performance and lower governance overhead.

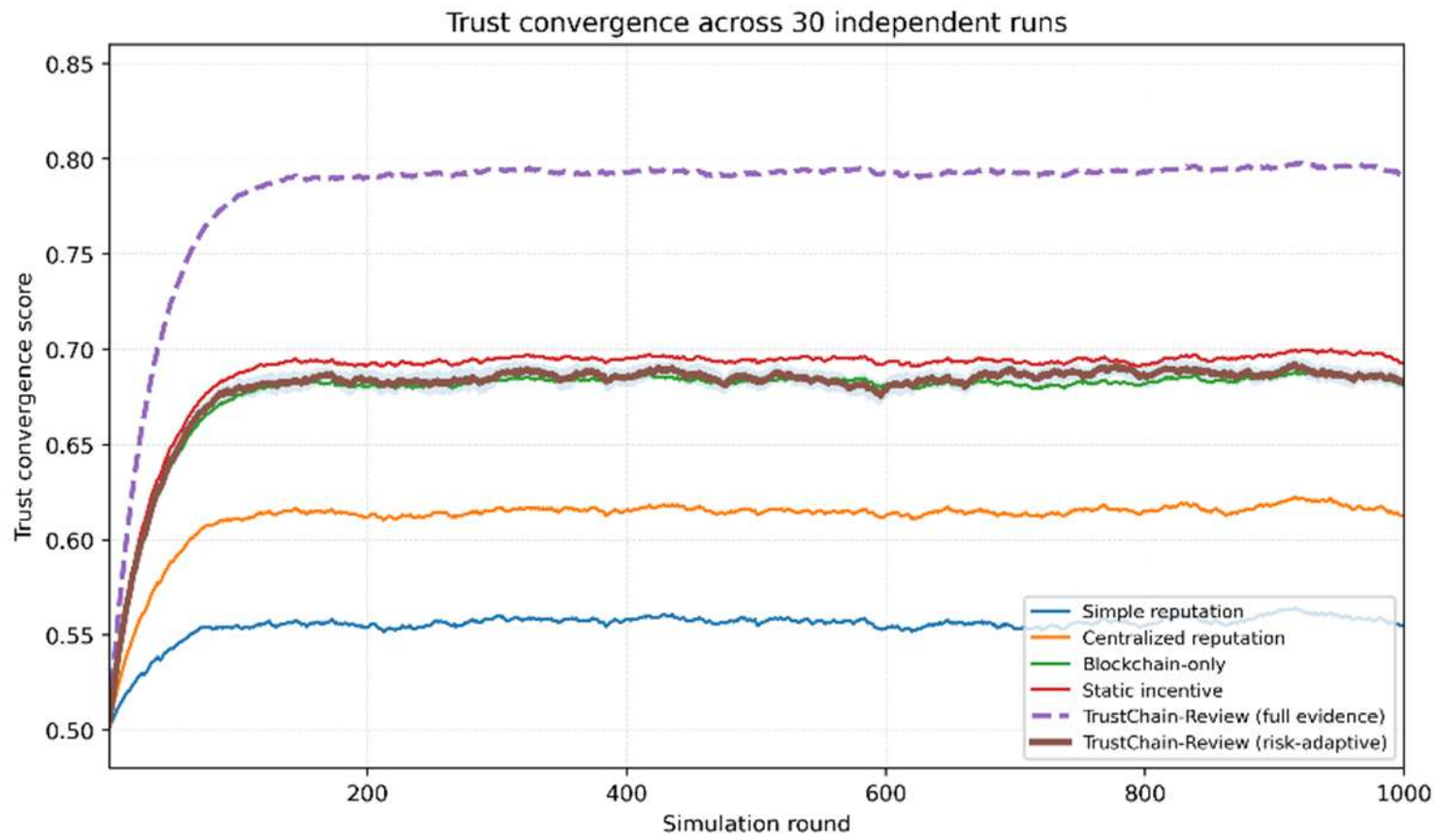

Fig. 1. Mean trust-convergence trajectories across 30 independent simulation runs.

Fig. 2 presents the malicious- and superficial-review detection results. Full-evidence TrustChain-Review achieves the highest rates, at $(0.790 \pm 0.035)$ and $(0.733 \pm 0.031)$, respectively. The risk-adaptive configuration achieves $(0.629 \pm 0.045)$ for malicious-review detection and $(0.608 \pm 0.040)$ for superficial-review detection. Its malicious-review detection is lower than that of the blockchain-only and static-incentive settings, whereas its superficial-review detection remains higher than that of simple reputation, centralized reputation, and blockchain-only governance. These results reflect the expected reduction in protection when evidence-based governance is not applied to every contribution.

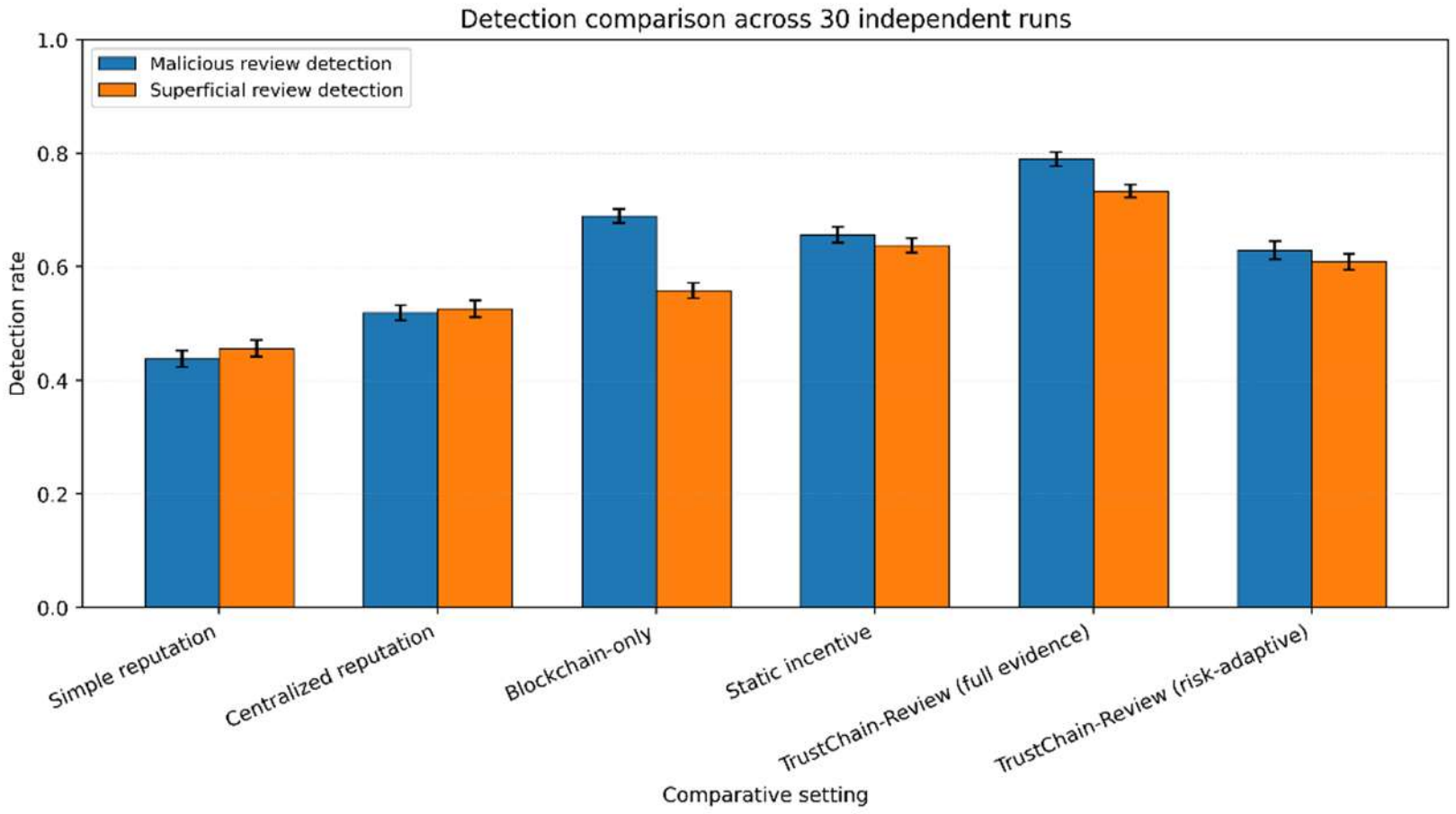

Fig. 2. Mean malicious and superficial review detection rates across 30 independent simulation runs; error bars indicate standard deviation.

To examine the sensitivity of the risk-triggered governance mechanism, the threshold $(r^*)$ was varied from 0.44 to 0.54 across 30 independent runs. As shown in Fig. 3, increasing the threshold reduces the proportion of contributions assigned to evidence-based governance and consequently lowers governance cost. At ( $r^*$=0.44), evidence-based governance is selected for approximately 93.6% of contributions, producing a mean governance cost of 0.315 and a cost-efficiency of 1.583. At the baseline threshold ($r^*$=0.50), the selection rate decreases to approximately 39.4%, governance cost decreases to 0.202, and cost-efficiency increases to 2.509. At ($r^*$=0.54), evidence-based governance is no longer selected, reducing the mechanism to a basic-governance configuration with a cost of 0.120 and a cost-efficiency of 3.102. This extreme setting is inexpensive but removes the additional trust and detection protection provided by evidence-based governance. Therefore, ($r^*$=0.50) is used as a transparent baseline compromise between governance coverage, operational cost, and cost-efficiency rather than being claimed as a universally optimal threshold.

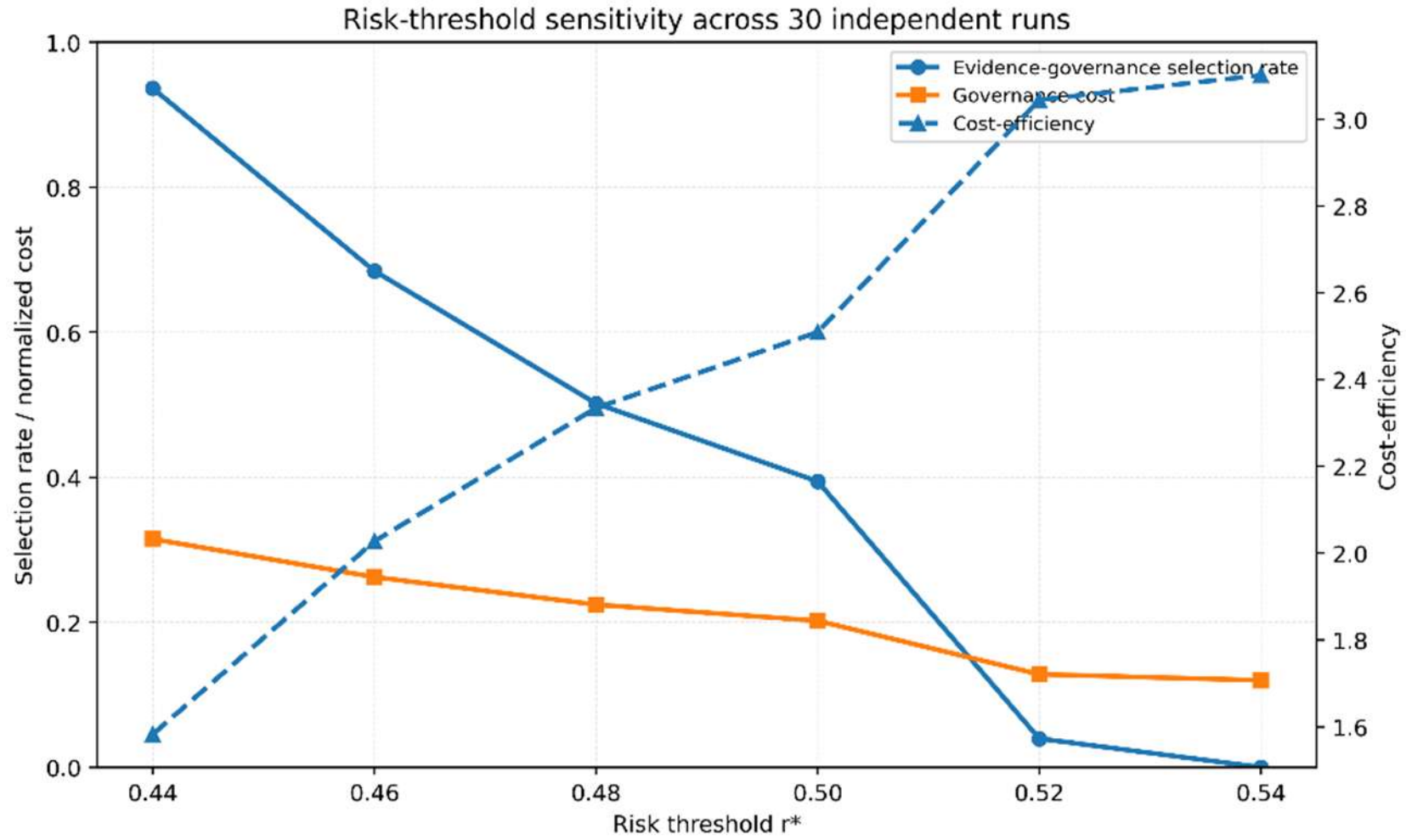

Fig. 3. Sensitivity of evidence-governance selection rate, governance cost, and cost-efficiency to the risk threshold ($r^*$) across 30 independent simulation runs.

# 9. Discussion and Research Implications

## 9.1. Interpretation of Empirical and Simulation Findings

The empirical variables support reproducible calibration but remain proxies rather than direct measures of semantic correctness, security severity, reviewer competence, or software criticality. Full-evidence governance provides the strongest trust and unreliable-review detection, whereas risk-adaptive governance lowers cost and improves efficiency by applying stronger controls selectively. No configuration therefore dominates every objective; governance choice depends on context.

## 9.2. Implications for Trustworthy AI-Assisted Code Review

AI-generated or modified code may appear plausible despite hidden defects [15], so trust should depend on the evidence supporting both contribution and review. TrustChain-Review separates developer and reviewer reliability architecturally, but the present simulation reports one aggregate reputation-accuracy metric; actor-level validation remains future work. Risk-triggered governance reserves stronger controls for higher-risk changes. Patch length is only a baseline proxy, and production systems should also consider criticality, affected subsystem, static-analysis and security findings, test coverage, and vulnerability indicators.

## 9.3. Implications for Blockchain-Based Evidence and Reputation

Blockchain strengthens traceability but cannot replace evidence assessment, reputation management, or behavioural incentives; the blockchain-only setting does not achieve the strongest overall performance [12], [13], [17]. TrustChain-Review records hashes, timestamps, identifiers, evidence references, and reputation-update summaries on-chain while keeping source code and complete comments off-chain. Evidence-based governance need not be uniform: full evidence maximizes protection, whereas risk-adaptive use limits overhead. Because recorded evidence may still be incomplete or inaccurate, blockchain logging must remain linked to accountability, reputation updates, incentives, and risk–cost decisions.

## 9.4. Practical Trade-Offs and Deployment Considerations

Full-evidence governance is most suitable for high-risk or audit-sensitive projects, while risk-adaptive governance better serves cost-sensitive settings. Thresholds should reflect project criticality, failure consequences, audit obligations, and available resources rather than be treated as universal constants.

## 10. Conclusion

This study presented TrustChain-Review, a blockchain-enabled, game-theoretic, and risk-adaptive framework for trustworthy AI-assisted code review. The framework connects auditable evidence, reputation, strategic incentives, and threshold-based selection between basic and evidence-based governance.

The empirical analysis of 328,340 dataset records showed that review-needed changes generally have larger patches and consistent comment availability. These observations calibrate the simulation, while patch length remains only a pre-review risk proxy.

Across 30 paired runs, full-evidence governance achieved the strongest reputation, trust, and unreliable-review detection but incurred the highest cost. Risk-adaptive governance reduced cost by approximately 38.4% and improved cost-efficiency by approximately 71.3%, while accepting lower trust and detection performance. The threshold therefore controls the trade-off between protection and overhead and should not be treated as universally optimal.

Future work should evaluate richer risk indicators, observed AI-assistance and post-merge outcomes, production blockchain performance, and live repository deployments.

**Code and Data Availability:** The complete reproducibility package for this study, including the risk-adaptive simulation code, configurations for 30 independent runs, generated simulation histories, statistical summaries, paired comparisons, threshold-sensitivity results, and publication-ready figures, is publicly available in Zenodo as version v1.1.0 at https://doi.org/10.5281/zenodo.20723944. The corresponding source repository is available at https://github.com/mohammadnaserameri/TrustChain-Review-Reproducibility. The original Diff Quality Estimation dataset from the CodeReviewer benchmark is not redistributed in the repository; users should obtain it from its original source, subject to the applicable access and licensing conditions.

**Funding :**This research received no specific grant from any funding agency in the public, commercial, or not-for-profit sectors.

**Declaration of Competing Interest :**The author declares that there are no known competing financial interests or personal relationships that could have appeared to influence the work reported in this paper.

**CRediT Authorship Contribution Statement:** Mohammad Naserameri: Conceptualization, Methodology, Formal analysis, Software, Validation, Investigation, Data curation, Writing – original draft, Writing – review & editing, Visualization.